\documentclass[11pt]{article}
\usepackage{amsmath}
\usepackage{amsfonts}

\title{
	Fuzzy de Sitter space-times via coherent states quantization}

\author{J-P. Gazeau\thanks{gazeau@ccr.jussieu.fr}, J. Mourad, and J. Queva\\ 
\emph{Boite 7020, APC, CNRS
	UMR 7164,}\\
	\emph{Universit\'{e} Paris 7-Denis Diderot, 75251 Paris Cedex 05,
	France}}
\begin{document}
\maketitle
\abstract{
	A construction of the 2d and 4d fuzzy de Sitter hyperboloids is carried out by using a (vector) coherent state quantization. We get a natural discretization of the dS ``time" axis based on the spectrum of Casimir operators of the respective maximal compact subgroups $SO(2)$ and $SO(4)$ of the de Sitter groups $SO_0(1,2)$ and $SO_0(1,4)$. The continuous  limit at infinite spins is examined. }

\section{Introduction}

	The Madore construction of the fuzzy sphere \cite{Madore} is based on the replacement  of coordinate functions of the sphere by components of the angular momentum operator in a $(2j+1)$-dimensional UIR of $SU(2)$. In this way, the commutative algebra of  functions on $S^2$, viewed as restrictions of smooth functions on $\mathbb{R}^3$, becomes the non-commutative algebra of $(2j+1)\times (2j+1)$-matrices, with corresponding differential calculus. The commutative limit is recovered at $j \to \infty$ while another parameter, say $\rho$,  goes to zero with the constraint $j\,\rho = 1$ (or $R$ for a sphere of radius $R$). The  aim of the present  work is to achieve a similar construction for the $2d$ and $4d$ de Sitter hyperboloids. The method is based on a generalization of coherent state quantization \emph{\`{a} la} Klauder-Berezin (see \cite{GPW,AEG} and references therein).  We recall that the de Sitter space-time is  the  unique maximally symmetric solution of the vacuum Einstein's equations with positive cosmological constant $\Lambda$. This constant is linked to the constant Ricci  curvature $4 \Lambda$ of this space-time. There exists a fundamental length $H^{-1} := \sqrt{3/(c\Lambda)}$. The isometry group of the de Sitter manifold is the ten-parameter de Sitter group $SO_0(1,4)$, the latter is a deformation of the proper orthochronous Poincar\'{e} group $\mathcal{P}_{+}^\uparrow$. 

\section{Coherent state quantization: the general framework}

	Let $X$  be a set equipped with the measure $\mu(dx)$ and $L^2(X, \mu)$ its associated Hilbert space of square integrable functions $f(x)$ on  $X$. Among the elements of $L^2(X, \mu)$ let us select an orthonormal set $\{ \phi_n(x),\  n = 1,2, \dotsc, N\}$, $N$ being finite or infinite, which spans, by definition,  a separable Hilbert subspace ${\mathcal H}$. This set is constrained to obey:  $0 < {\cal N} (x) := \sum_n \; \lvert \phi_n (x) \rvert^2 < \infty.$
Let us then consider the family of states $\{\lvert x \rangle \}_{x\in X}$ in $ \mathcal{H}$ through the following linear superposition: 
\begin{equation}
	\label{CS}
	\lvert x\rangle := \frac{1}{\sqrt{\mathcal{N} (x)}} \sum_n \overline{\phi_n (x)} \lvert\phi_n\rangle  .
\end{equation}
This defines an injective map (which should be continuous w.r.t.  some topology affected to $X$) $ X \ni x\mapsto \lvert x \rangle \in \mathcal{H}$. These coherent states
are normalized and provide a resolution of the unity in $\mathcal{H}$:
\begin{equation}
	\langle x \,|\, x \rangle = 1, \ \
	\int_X | x\rangle \langle x  | \, {\mathcal N}(x)\,\mu(dx)= \mathbb{I}_{{\mathcal H}}. \label{normaresun}
\end{equation}
A {\it classical} observable is  a function $f(x)$ on $X$ having specific properties. Its  quantization   \emph{\`a la} Berezin-Klauder-``Toeplitz''  consists in associating to $f(x)$ the operator
\begin{equation}
	\label{quantber}
	A_f := \int_X f(x) \lvert x\rangle \langle x\rvert \, {\mathcal N}(x)\,\mu(dx).
\end{equation}
For instance, the application to the sphere $X = S^2$ with normalized measure $\mu (dx) = \sin \theta ~d\theta~d\phi/4\pi$ is carried out through the choice as orthonormal set  the set of \emph{spin spherical harmonics} $_{\sigma}Y_{jm}(\hat{\mathbf{r}})$ for fixed $\sigma$ and $j$. One obtains \cite{gahulare} in this way a  family of inequivalent (with respect to quantization) fuzzy spheres, labeled by the  the spin parameter  $0 < \lvert \sigma\rvert  \leq j, \, j \in \mathbb{N}_{\ast}/2$. Note that the spin is necessary in order to get a nontrivial quantization of the cartesian coordinates.

\section{Application to the $2d$  de Sitter space-time}
 
De Sitter space is seen as a one-sheeted hyperboloid embedded in a three-dimensional Minkowski  space:
\begin{equation}
	\label{contrainte-geometrie}
	M_H = \{x \in \mathbb{R}^3 \;:\; x^2 = \eta_{\alpha\beta}\,x^\alpha x^\beta = (x^0)^2 - (x^1)^2 - (x^2)^2 = -H^{-2}\}. 
\end{equation}
	The de Sitter group is $SO_0(1,2)$ or its double covering $SU(1,1) \simeq SL(2,\mathbb{R})$. Its Lie algebra is spanned by the three Killing vectors $K_{\alpha \beta} = x_{\alpha}\partial_{\beta} - x_{\beta}\partial_{\alpha} $ 
($K_{12}$: compact, for ``space translations'', $K_{02}$: non compact, for ``time translations'', $K_{01}$: non compact, for Lorentz boosts).
These Killing vectors are represented as (essentially) self-adjoint operators in a Hilbert space of  functions on $M_H$, square integrable with respect to some invariant inner (Klein-Gordon type) product.

The quadratic Casimir operator has  eigenvalues  which  determine  the UIR's :
\begin{equation}
 \label{contrainte-groupe}
Q = - \frac{1}{2} M_{\alpha\beta}M^{\alpha\beta} = - j(j+1)\mathbb{I} = \Big(\rho^2 + \frac{1}{4}\Big)\mathbb{I}
\end{equation}
where $j = -\frac{1}{2} + i \rho$, $\rho\in\mathbb{R}^{+}$ for the principal series.

	Comparing the geometric constraint \eqref{contrainte-geometrie} to the group theoretical one \eqref{contrainte-groupe} (in the principal series) suggests the fuzzy correspondence \cite{gamouque}:
\begin{equation*}
 x^\alpha \mapsto \widehat{x^\alpha} = \frac{r}{2}\varepsilon^{\alpha\beta\gamma}\,M_{\beta\gamma}, \; \mbox{{\em i.e.}} \;\;
 \widehat{x^0} = r M_{21}, \ \widehat{x^1} = r M_{02}, \ \widehat{x^2} = r M_{10}.
\end{equation*}
$r$ being a constant with  length dimension. The following commutation rules are expected 
\begin{equation}
 \label{comm-2d}
	[\widehat{x^0},\, \widehat{x^1}] = i r \widehat{x^2}, \;
	[\widehat{x^0},\, \widehat{x^2}] = - i r \widehat{x^1},\; 
	[\widehat{x^1},\, \widehat{x^2}] = i r \widehat{x^0},
\end{equation}
with $\eta_{\alpha\beta} \widehat{x^\alpha}\widehat{x^\beta} = - r^2(\rho^2 + \frac{1}{4})\mathbb{I}$, and its ``commutative classical limit'', $r \to 0, \, \rho \to \infty, \, r \rho = H^{-1}$.

	Let us now  proceed to the CS quantization of the 2d dS hyperboloid. The ``observation'' set $X$ is the hyperboloid $M_{H}$. Convenient global coordinates are those of the topologically equivalent cylindrical structure:
 $(\tau, \theta),\ \tau \in \mathbb{R}, \ 0\leq \theta < 2 \pi, $
through the parametrization, $x^0 = r\tau, \ x^1 = r \tau\cos{\theta} - H^{-1} \sin{\theta}, \ x^2 = r\tau\sin{\theta} + H^{-1}\cos{\theta}$, with the invariant measure: $\mu(dx) = \frac{1}{2\pi}  \, d\tau\, d\theta $. The functions $\phi_m (x)$ forming the orthonormal system needed to construct coherent states are suitably weighted Fourier exponentials:
\begin{equation}
	\label{gauss}
	\phi_m (x) = \left(\frac{\epsilon}{\pi}\right)^{1/4}\,e^{-\frac{\epsilon}{2}(\tau-m)^2 } \,e^{ i m\theta}, \ m\in \mathbb{Z} ,
\end{equation}
where the parameter  $\epsilon > 0$ can be arbitrarily small and represents a necessary regularization. Through the superposition \eqref{CS} the coherent states read
\begin{equation}
	\label{ccs}
	\lvert \tau , \theta \rangle =  \frac{1}{\sqrt{\mathcal{N} (\tau)}}\,  \left(\frac{\epsilon}{\pi}\right)^{1/4} 
		\sum_{m \in \mathbb{Z}} e^{-\frac{\epsilon}{2}(\tau-m)^2 } \,e^{- i m\theta}  \lvert m \rangle,
\end{equation}
where $\lvert \phi_m\rangle \simeq \lvert m \rangle.$ The normalization factor $\mathcal{N}(\tau) = \sqrt{\frac{\epsilon}{\pi}}\sum_{m \in \mathbb{Z}} e^{-\epsilon (\tau-m)^2} < \infty$ is  a periodic train of normalized Gaussians and is proportional to an elliptic Theta function. 
 
	The CS quantization scheme \eqref{quantber} yields the quantum operator $A_f$, acting on $\mathcal{H}$ and associated to the classical observable $f(x)$. For the most basic one, associated to the coordinate $\tau$, one gets
\begin{equation}  
	A_{\tau}  
		= \int_{X} \tau\, \lvert \tau,\theta \rangle\langle \tau,\theta\rvert  \mathcal{N}(\tau)\,\mu (dx)  
		= \sum_{m \in \mathbb{Z}} m  \lvert m\rangle \langle m\rvert .
\end{equation}
This  operator  reads in angular position representation (Fourier series): $A_{\tau} = -i\frac{\partial}{\partial \theta}$,
and is easily identified as the compact representative $M_{12}$ of the Killing vector $K_{12}$ in the principal series UIR. Thus, the ``time'' component $x^0$ is naturally quantized, with spectrum $r\mathbb{Z}$ through $x^{0} \mapsto \widehat{x^0} = -r M_{12}$. For the two other ambient coordinates one gets:
\begin{equation*}
	\widehat{x^1} = \frac{r e^{-\frac{\epsilon}{4}}}{2}\sum_{m\in\mathbb{Z}}
		\big\{p_m |m+1\rangle\langle m| + h.c \big\}\,,\;
	\widehat{x^2} = \frac{r e^{-\frac{\epsilon}{4}}}{2i}\sum_{m\in\mathbb{Z}}
		\big\{p_m |m+1\rangle\langle m| - h.c \big\},
\end{equation*}
 with $p_m = (m + \frac{1}{2} + i\rho)$. Commutation rules are those of $so(1,2)$, that is those of \eqref{comm-2d} with a local modification to $[\widehat{x^1},\, \widehat{x^2}] = -i re^{-\frac{\epsilon}{2}}\widehat{x^0}$. The commutative limit at $r \to 0$ is apparent. It is proved that the same holds for higher degree polynomials in the ambient space coordinates.

\section{Application to the 4d de Sitter space-time}
 
	The extension of the method to the 4d-de Sitter geometry and kinematics involves the universal covering of $SO_0(1,4)$, namely, the symplectic 
$Sp(2,2)$ group, needed for half-integer spins. In a given UIR of the latter,  the ten Killing vectors  are represented as (essentially) self-adjoint operators in  Hilbert space of (spinor-)tensor valued functions on the de Sitter space-time $M_H$, square integrable with respect to some invariant inner (Klein-Gordon type) product : $K_{\alpha\beta} \rightarrow L_{\alpha\beta}$. There are now two Casimir operators whose eigenvalues   determine  the UIR's:
\begin{equation*}
	Q^{(1)} = - \frac{1}{2} L_{\alpha\beta}L^{\alpha\beta}, \
	Q^{(2)} = - W_{\alpha} W^{\alpha}, \ W^\alpha := -\frac{1}{8}\epsilon^{\alpha\beta\gamma\delta\eta} L_{\beta\gamma}L_{\delta\eta}.
\end{equation*}
Similarly to the 2-dimensional case, the principal series  is involved in the construction of the fuzzy de Sitter space-time. Indeed, by comparing both constraints, the geometric one: $ \eta_{\alpha\beta} x^{\alpha}x^{\beta} = -H^{-2}$ and the group theoretical one, involving the \emph{ quartic} Casimir (in the principal series with  spin $s >0$): 
 $Q^{(2)} = - W^{\alpha} W_{\alpha}= \left( \nu^2 +  \frac{1}{4}\right) \,s(s+1)\,\mathbb{I}$
suggests the correspondence \cite{gamouque}: $x^{\alpha} \mapsto \widehat{x^{\alpha}} = r W^{\alpha}$, and the ``commutative classical limit'' :  $r \to 0, \nu \to \infty, \, r s \sqrt{\nu^2 + \frac{1}{4}}  = H^{-1}$.

	For the CS quantization of the 4d-dS hyperboloid, suitable global coordinates are those of the topologically equivalent  $\mathbb{R} \times S^3$ structure:  $(\tau, \xi),\ \tau \in \mathbb{R}, \ \xi \in S^3$, through the following parametrization, $x^0 = r\tau, \ \mathbf{x}= (x^1,x^2,x^3,x^4)^{\dag} = r \tau \,\xi  + H^{-1}\, \xi^{\bot},$ where $\xi^{\bot} \in S^3$ and $\xi\cdot \xi^{\bot} = 0$, with the invariant measure: $\mu(dx) = d\tau\, \mu(d\xi)$.
We now consider the  spectrum $\{ \tau_i \, \mid \, i \in \mathbb{Z}\}$ 
 of the compact ``dS fuzzy time'' operator $ r W^0$
 in the Hilbert space  $ L^2_{\mathbb{C}^{2s+1}}(S^3)$ which carries the principal series UIR $U_{s, \nu}$, $s > 0$. 
 This spectrum is discrete. Let us denote by  $\{\mathcal{Z}_{\mathcal{J}} (\xi)\}$, where $\mathcal{J}$ represents a set of  indices including in some way the index $i$, an orthonormal basis 
of  $ L^2_{\mathbb{C}^{2s+1}}(S^3)$ made up with the eigenvectors of $W^0$. 
The functions $\phi_{\mathcal{J}} (x)$, forming the orthonormal system needed to construct coherent states, are suitably weighted Fourier exponentials:
\begin{equation}
	\label{1+3os}
	\phi_{\mathcal{J}}(x) = \left(\frac{\epsilon}{\pi}\right)^{1/4}\,e^{-\frac{\epsilon}{2}(\tau- \tau_i)^2 } \, \mathcal{Z}_{\mathcal{J}} (\xi) ,
\end{equation}
where $\epsilon > 0$ can be arbitrarily small. 
The resulting vector coherent states read as
\begin{equation}
	\label{1+3ccs}
	\lvert \tau , \xi \rangle =  \frac{1}{\sqrt{{\mathcal N} (\tau, \xi)}}\,  \left(\frac{\epsilon}{\pi}\right)^{1/4} 
		\sum_{\mathcal{J}} e^{-\frac{\epsilon}{2}(\tau-\tau_i)^2 } \,  \overline{\mathcal{Z}_{\mathcal{J}} (\xi)} \lvert \mathcal{J}\rangle,
\end{equation}
with normalization factor 
\begin{equation*}
	\mathcal{N}(x) \equiv \mathcal{N}(\tau, \xi) 
		= \sqrt{\frac{\epsilon}{\pi}}\sum_{\mathcal{J}} e^{-\epsilon (\tau-\tau_i)^2} \mathcal{Z}^{\dag}_{\mathcal{J}} (\xi) \mathcal{Z}_{\mathcal{J}} (\xi)< \infty.
\end{equation*}

\end{document}